\documentclass[preprint,12pt,authoryear]{elsarticle}

\usepackage{amssymb}
\usepackage{natbib}
\usepackage{amsfonts}
\usepackage{amsmath}
\usepackage{amscd}
\usepackage{subfigure}
\usepackage{color}
\usepackage{amsthm}
\usepackage{graphicx}
\usepackage{epstopdf}
\usepackage{multirow}

\def\zb{{\bf z}}

\def\yb{{\bf y}}
\def\Wb{{\bf W}}
\def\Ab{{\bf A}}
\def\Bb{{\bf B}}

\def\Db{{\bf D}}
\def\Ib{{\bf I}}

\def\Pb{{\bf P}}
\def\1b{{\bf 1}}

\journal{arXiv.org}

\begin{document}

\begin{frontmatter}



\title{Measuring the spatial balance of a sample:\\
A new measure based on the Moran's $I$ index}


\author[label1]{Yves Till\'e}
\author[label2]{Maria Michela Dickson}
\author[label2]{Giuseppe Espa}
\author[label2]{Diego Giuliani}
\address[label1]{Institut de Statistique, Universit\'e de Neuch\^atel - Switzerland}
\address[label2]{Department of Economics and Management, University of Trento - Italy}

\author{}

\address{}

\begin{abstract}
Measuring the degree of spatial spreading of a sample can be of great interest when sampling from a spatial population. The commonly used spatial balance index by \citet{LPM2011} is particularly effective in comparing the level of spatial spreading of different samples from the same population. However, its unbounded and uninterpretable scale of measurement does not allow to assess the level of spatial spreading in absolute terms and confines its use to only raw comparisons. In this paper, we introduce a new absolute measure of the spatial spreading of a sample using a normalized version of the Moran's $I$ index. The properties and behaviour of the proposed measure are analysed through two simulation experiments, one based on artificial populations and the
other on a population of real business units located in the province of Siena (Italy).

\end{abstract}

\begin{keyword}
Spatial sampling \sep Spatial balance \sep Moran's $I$ index \sep Statistical measure
%% keywords here, in the form: keyword \seep keyword

%% PACS codes here, in the form: \PACS code \seep code

%% MSC codes here, in the form: \MSC code \seep code
%% or \MSC[2008] code \seep code (2000 is the default)

\end{keyword}

\end{frontmatter}

%% \line numbers

\section{Introduction}

Since the seminal paper by \citet{StevenOlsen2004}, spatial balance is considered as one of the most important and useful property of a sample selected from a spatial population. A recent large body of literature on survey methodology has shown that when the target variable is characterized by some form of spatial autocorrelation or spatial heterogeneity, the samples with units that are well spread in space, or say the spatially balanced samples, tend to provide relatively more efficient estimates of the population mean or total \cite[see, among others,][]{LPM2011,SCPS2012,GrafTille2013}.

The main formal measure of the degree of spatial balance for a sample is the one proposed by \citet{LPM2011}, based on the approach of Voronoi polygons first suggested by \citet{StevenOlsen2004}. Consider a finite two-dimensional spatial population $U$ of $N$ units, and assume that each unit $i$ has a given inclusion probability $\pi_i$ such that $\sum_{i=1}^{N} \pi_i = n$, where $n$ is the fixed sample size. For a sample  $S = (s_1,s_2,\dots,s_n)$ selected from $U$, the Voronoi polygon associated to the sample unit $s_i \in S$ is the set of all units of population $U$ that are closer to $s_i$ than to any other sample unit $s_j$. Moreover, let $v_i$ denote the sum of all the inclusion probabilities of units belonging to the $i$th polygon. According to \citet{LPM2011}, a sample is perfectly spatially balanced if all the $v_is$ are equal to 1. Therefore, \citet{LPM2011} suggest that the variance
$$
B(S) = \frac{1}{n} \sum_{i=1}^{n} (v_i-1)^2
$$
can represent a proper measure of spatial balance for a sample $S$. The lower its value is, the higher is the degree of spatial balance of $S$, that is, the better spread in space are the units of sample $S$.

The feasible range of $B$ depends on the spatial pattern of the population $U$ and hence differs among different populations, which implies that $B$ can be used successfully to compare the level of spatial balance of different samples from the same population. Because of that, this measure has recently become the standard tool to assess which sampling design is better than others at selecting well spread samples \cite[see, for example,][]{CEUS2017,Tille2017}. Nevertheless, as a spatial balance index, $B$ has some inherent limitations that restrict its applicability to only raw comparisons. In particular, $B$ lacks of a useful interpretation, does not vary within a fixed finite range and is not characterized by a specific benchmark value that discriminates between absence and presence of spatial balance.

In this paper, we introduce a new spatial balance index that overcomes the above limitations. The new index has a finite meaningful range, from -1 (perfect spatial balance) to +1 (maximum concentration), and a clear-cut benchmark value, namely 0. These properties make the index able to evaluate meaningfully the degree of spatial spreading of a sample and assess whether it is spatially balanced or spatially clustered. Therefore, the proposed measure can have a more general applicability than just assessing which spatial sampling design is better at spreading for a given population. For example, if the aim of a survey is to collect data in order to estimate spatial autocorrelation or to detect small-scale spatial structures, a sample characterized by spatial concentration may be preferable than a well spread sample. Unlike the $B$ index, the new measure, which essentially consists on an aptly modified version of the Moran's $I$ index of spatial autocorrelation \cite[]{Moran1950}, can effectively indicates whether the selected sample has a proper level of spatial concentration.

The structure of the paper is the following. Section~2 reviews the traditional Moran's $I$ index of spatial autocorrelation and shows how it can be used to measure the spatial spreading of a sample. Section~3 introduces the new spatial balance measure, $I_B$, which is based on a specific normalized version of the Moran's $I$ index. Section~4 describes how to properly specify the spatial weights matrix for $I_B$. Section~5 analyses the behaviour and properties of $I_B$ through a simulation experiment based on artificial spatial populations. The same analysis is then performed in Section~6 with an population of actual business units. Finally, Section~7 gives an account of some general conclusions.

\section{Measuring the spatial spreading of a sample through the traditional Moran's $I$ index}

The proposed approach to measure the degree of spatial balance of a sample is based on the simple intuition that the level of spatial spreading of the sample units is reflected by the level of spatial autocorrelation of the sample inclusion indicator variable. Let $U = (1, 2, \dots, i, \dots, N)$ be the finite spatial population of size $N$, and $S$ be a random sample drawn from $U$. The sample inclusion indicator variable $\delta_i$, observed for the population unit $i$, specifies whether $i$ is included in $S$ or not, that is
$$
\delta_i = \begin{cases} 1 & \mbox{ if unit } i \in S \\ 0 & \mbox{ otherwise.} \end{cases}
$$

\begin{figure}[htb!]
\begin{center}
\caption{A stylized representation of the relationship between the spatial spreading of a sample and the spatial distribution of the sample inclusion indicator. Map on the left depicts a well spread sample and a negatively autocorrelated inclusion indicator, map on the right depicts a spatially clustered sample and a positively autocorrelated inclusion indicator.}
\includegraphics[scale=0.35]{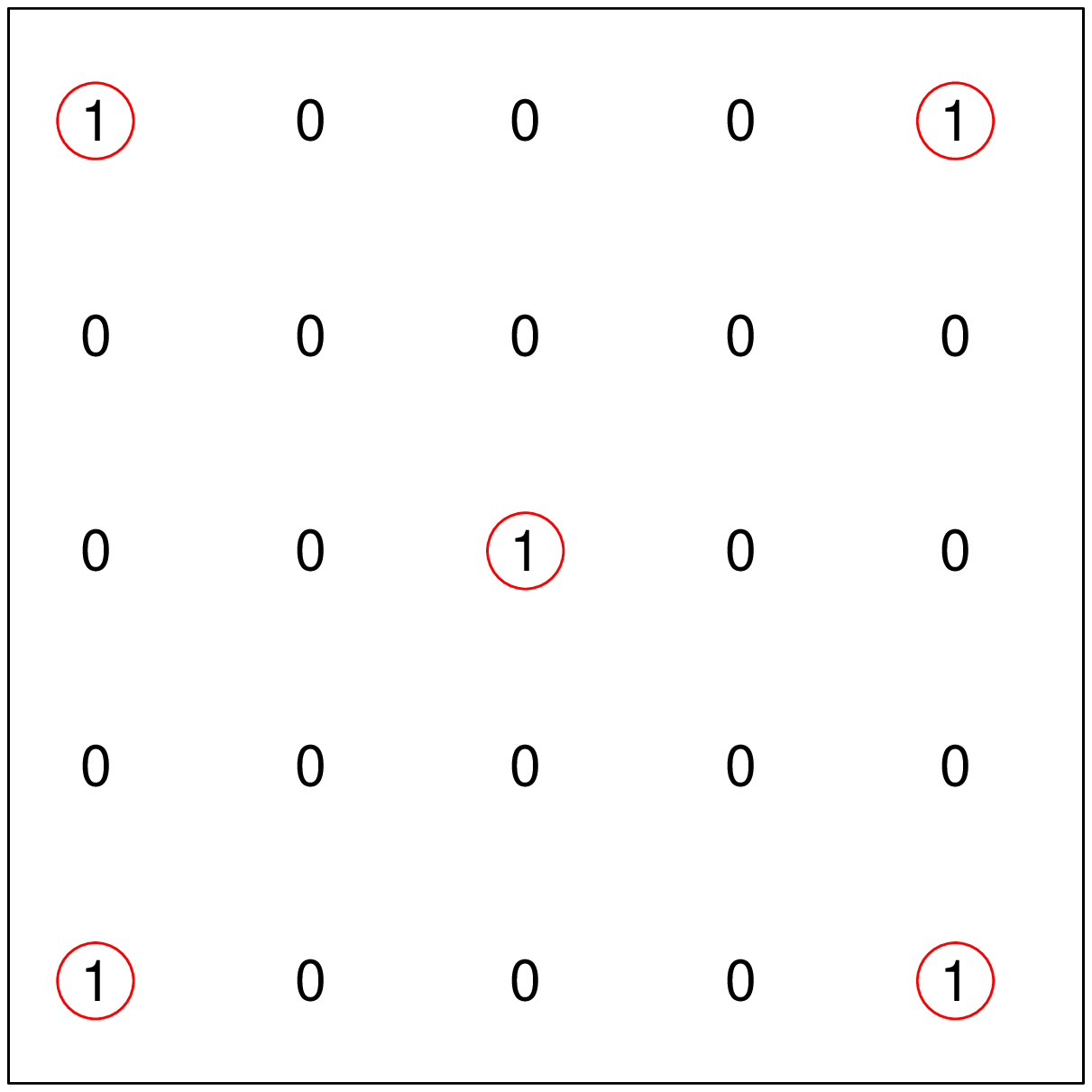}\includegraphics[scale=0.35]{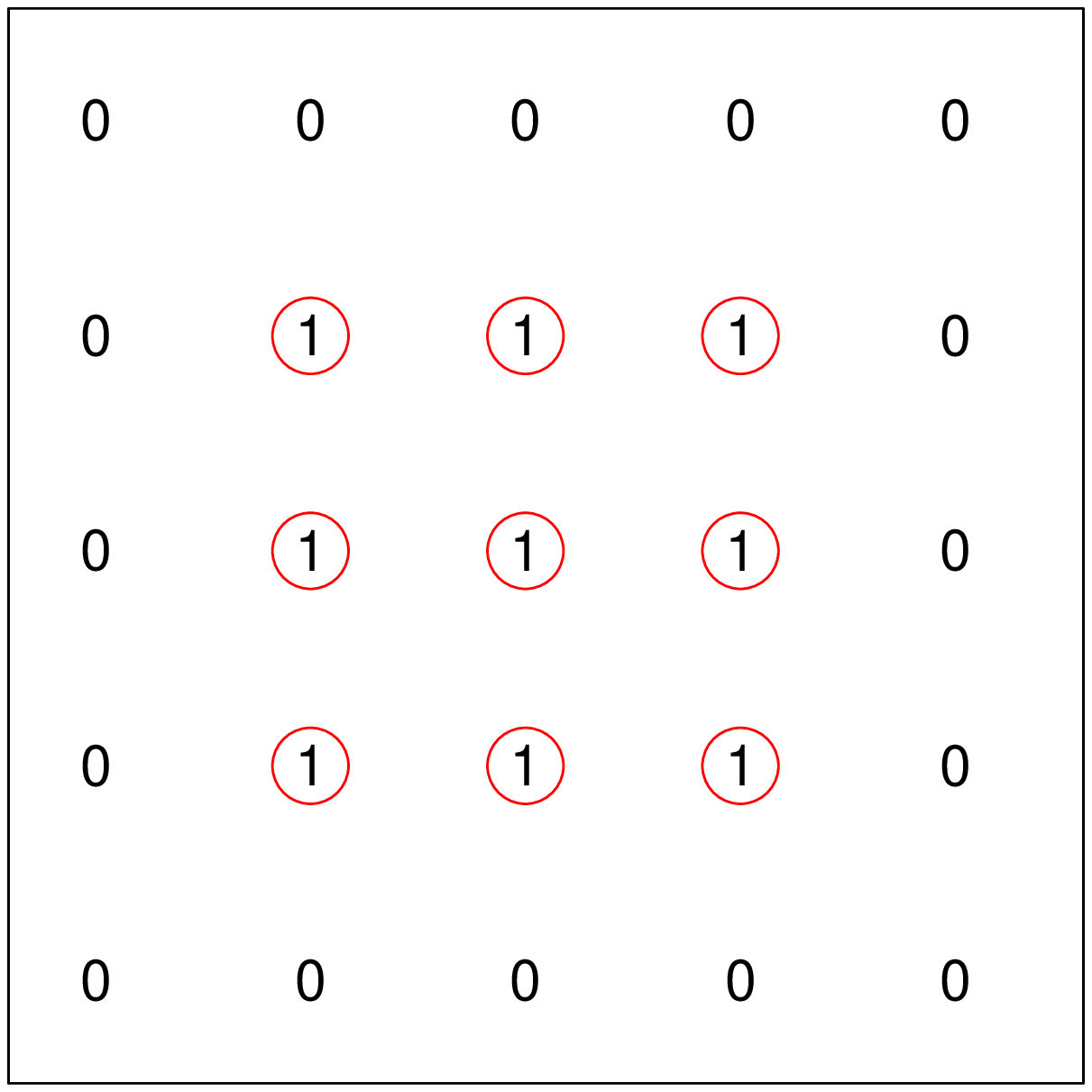}
\label{fig:graphInclusionIndicator}
\end{center}
\end{figure}

As shown in the stylized illustration of Figure~\ref{fig:graphInclusionIndicator}, when the sample is well spread in space, the sample units tend to be located relatively far apart from each other and hence variable $\delta$ is negatively autocorrelated. On the other hand, if the sample is spatially clustered, the sample units tend to be located relatively closely to each other and therefore $\delta$ is positively autocorrelated.

According to this perspective, a measure of the degree of spatial autocorrelation for $\delta$ can function as a measure of spatial spreading of the sample. A natural candidate to measure the degree of spatial autocorrelation of a single variable is the popular Moran's $I$ index \cite[]{Moran1950}. This index has the desired properties mentioned above. It varies in finite range and has a definite benchmark value distinguishing between negative and positive spatial autocorrelation. Unfortunately, as we will discuss below, although by construction the traditional Moran's $I$ index should vary in the fixed $[-1,+1]$ interval, this occurs only under limited empirical circumstances, implying that the interval may change according to the population. Therefore, we introduce a normalized version of the index that ensures a unique fixed variation range.

Let $y_i$ be the value taken by variable $y$ on unit $i$ and
$$
\overline{Y} = \frac{1}{N}\sum_{i\in U}y_i.
$$
The Moran's $I$ index is a global measure of spatial autocorrelation, for variable $y$, defined by
\begin{equation}
I = \frac{
N \sum_{i\in U}\sum_{j\in U}  w_{ij} (y_i-\overline{Y})(y_j-\overline{Y})
 }{
\left(\sum_{i\in U}\sum_{j\in U}  w_{ij} \right)\sum_{i\in U} (y_i - \overline{Y})^2
}
\label{moran}
\end{equation}
\cite[]{Moran1950}. The weights $w_{ij}$ are elements of a matrix of nonnegative spatial weights such that $w_{ii}=0$ for all $i\in U$.
In particular, $w_{ij}$ indicates how close is $j$ to $i$, and hence a large value of it means that $j$ is a neighbour of $i$. Matrix $\Wb=(w_{ij})$ is not supposed to be symmetric. Moreover, matrix $\Wb$ is also not supposed to be stochastic i.e. the sum of the rows are not necessarily equal to one.

Let $\yb$ be  the vector of $y_i$, $\1b$ be a vector of $N$ ones, $\bar{\yb}=\overline{Y}\1b=  \1b\1b^\top\yb /N,$
and
$\Pb= \Ib-\1b\1b^\top\yb /N$.
Expression~(\ref{moran}) can therefore also be written as
\begin{equation}
I = \frac{(\yb-\bar{\yb})^\top \Wb(\yb-\bar{\yb})}{(\yb-\bar{\yb})^\top(\yb-\bar{\yb}) \;\; \1b^\top \Wb \1b}
= \frac{ \yb^\top \Pb \Wb\Pb \yb}{\yb^\top\Pb \yb  \;\; \1b^\top \Wb \1b} .
\label{moran2}
\end{equation}

The Moran's $I$ index is often presented as behaving like a correlation coefficient, ranging from $-1$ (for perfect negative spatial association) through approximately zero ($-1/(n-1)$ to be more precise), for spatial independence, to $1$ for perfect positive spatial association. Nevertheless, it does not work exactly as a correlation coefficient and, in particular, is not necessarily in the $[-1,+1]$ interval. A coefficient of correlation is a covariance divided by the two corresponding standard deviations. This is not the case of the Moran's $I$ index. An analysis of its extreme values shows that their absolute value can be larger or smaller than one
\cite[]{jong1984extreme,dray2011new}. Besides, in the original paper \cite[]{Moran1950}, the index is only defined for the particular case where the neighbours are the four contiguous units on a regular grid. It is not really surprising that the first normalization proposed by Moran cannot be generalized for any matrix of $w_{ij}.$

\section{Measuring the spatial spreading of a sample through a normalized Moran's $I$ index: the $I_B$ index of spatial balance}

In order to have an index that varies in the $[-1,+1]$ interval, we propose an alternative normalization of the Moran index.
Let
$$
w_{i\cdot} = \sum_{j\in U} w_{ij},
w_{\cdot j} = \sum_{i\in U} w_{ij}, \mbox{ and }
w= \sum_{i\in U}  \sum_{j\in U} w_{ij}\;\;.
$$

A useful simple way of defining a normalized index consists on computing a weighted correlation between $y_i$ and the local mean of the units in the neighbourhood of~$i$.
Let define
$$
\overline{Y}_w = \frac{1}{w}\sum_{i\in U}w_{i\cdot}y_i,
%\sigma^2_w= \frac{1}{w}\sum_{i\in U}w_{i\cdot}(y_i-\overline{Y}_w)^2,
$$
$$
z_i=y_i-\overline{Y}_w,\;\;\;
\overline{Z}_i = \frac{\sum_{j\in U} w_{ij}z_j}{\sum_{j\in U} w_{ij}}= \frac{\sum_{j\in U} w_{ij}z_j}{w_{i\cdot}}
$$
$$
\overline{\overline{Z}} = \frac{1}{w}\sum_{i\in U} w_{i\cdot}\overline{Z}_i = \frac{1}{w}\sum_{i\in U} w_{i\cdot} \frac{\sum_{j\in U} w_{ij}z_j}{\sum_{j\in U} w_{ij}}=
\frac{1}{w} \sum_{j\in U} w_{\cdot j}z_j .
$$
$$
\overline{Z}_i - \overline{\overline{Z}}  =     \sum_{j\in U} \left(\frac{w_{ij}}{w_{i\cdot}} -  \frac{w_{\cdot j}}{w}  \right) z_j
=     \sum_{j\in U} a_{ij} z_j,
$$
where
$$
a_{ij}=\frac{w_{ij}}{w_{i\cdot}} -  \frac{w_{\cdot j}}{w}.
$$

The local mean $\overline{Z}_i$ can be interpreted as the weighted average of the values for the units that are close to unit $i$. We can therefore define a normalized index between the values $z_i$ and the local mean $\overline{Z}_i$ by
\begin{eqnarray}
\tilde{I} &=& \frac{
\sum_{i\in U} w_{i\cdot} z_i (\overline{Z}_i-\overline{\overline{Z}})
}{\sqrt{
\sum_{i\in U} w_{i\cdot} z_i^2\;\;\;
\sum_{i\in U} w_{i\cdot} (\overline{Z}_i-\overline{\overline{Z}})^2
}
}
\nonumber\\
&=&\frac{\sum_{i\in U}w_{i\cdot} z_i \; \overline{Z}_i }
{\sqrt{\sum_{i\in U}w_{i\cdot} z_i^2 \sum_{i\in U}\;w_{i\cdot}\left(
\sum_{j\in U} a_{ij} z_j
\right)^2}}\nonumber\\
&=&\frac{\sum_{i\in U} \sum_{j\in U} z_i w_{ij}z_j }
{
\sqrt{
\sum_{i\in U} w_{i\cdot} z_i^2\;\;\;
\sum_{j\in U} z_j \sum_{k\in U} z_k \sum_{i\in U}w_{i\cdot}a_{ij}a_{ik}}
} \nonumber\\
&=&\frac{\sum_{i\in U} \sum_{j\in U} z_i w_{ij}z_j }
{\sqrt{
\sum_{i\in U} w_{i\cdot} z_i^2\;\;\;
\sum_{j\in U} z_j \sum_{k\in U} z_k b_{jk}} }\label{brol},
\end{eqnarray}
where
$$
b_{jk}=\sum_{i\in U}w_{i\cdot}a_{ij}a_{ik}.
$$

Index $\tilde{I}$ is thus a weighted correlation between $y_i$ and the average value of its neighbours.

Let $\zb$ be the vector of $z_i$ and $\Ab$, $\Bb$ be matrices of $a_{ij}$ and $b_{ij}$.
Let also  $\Db$ be the diagonal matrix containing $w_{i.}$ on its diagonal.
Expression~(\ref{brol}) can then also be written as
$$
\tilde{I} = \frac{\zb^\top \Wb\zb}{\sqrt{\zb^\top \Db\zb \;\;\;\; \zb^\top \Bb\zb}},
$$
where
$$
\Ab = \Db^{-1}\Wb - \frac{\1b\1b^\top \Wb }{\1b^\top \Wb \1b}
$$
$$
\Bb = \Ab^\top \Db \Ab = \Wb^\top \Db^{-1}\Wb
 - \frac{ \Wb^\top \1b \1b^\top \Wb }{\1b^\top \Wb \1b}.
 $$
 Note that $\Bb\1b={\bf 0}$ and that
 $$
 \overline{Y}_w = \frac{\yb^\top \Wb \1b}{\1b^\top \Wb \1b}.
 $$
 %$$
% \sigma^2_w=\frac{1}{w}(\yb-\bar{\yb})^\top \Db (\yb-\bar{\yb}).
% $$

If $\bar{\yb}_w=\1b  \overline{Y}_w$, the index can also be written as

\begin{eqnarray}
\tilde{I}= \frac{(\yb-\bar{\yb}_w)^\top \Wb (\yb-\bar{\yb}_w)}{\sqrt{ (\yb-\bar{\yb}_w)^\top \Db (\yb-\bar{\yb}_w) \;\;\;   (\yb-\bar{\yb}_w)^\top \Bb (\yb-\bar{\yb}_w)}}.
\label{Itilde}
\end{eqnarray}

When the sum of each row of matrix $\Wb$ is constant, $\bar{\yb}_w=\bar{\yb}.$ In this case, Expression~(\ref{Itilde}) has exactly the same numerator as the usual Moran's $I$ index, but the normalization is different.
The new index $\tilde{I}$ is a correlation coefficient between the $y$-values and their local means and is thus always in the $[-1,+1]$ interval.
If we compute $\tilde{I}$ for the sample inclusion indicator variable, that is if we substitute $\yb$ with $\boldsymbol\delta$ in Expression~(\ref{Itilde}), we obtain the desired spatial balance measure, that is
\begin{eqnarray}
I_B = \frac{(\boldsymbol\delta-\bar{\boldsymbol\delta}_w)^\top \Wb (\boldsymbol\delta-\bar{\boldsymbol\delta}_w)}{\sqrt{ (\boldsymbol\delta-\bar{\boldsymbol\delta}_w)^\top \Db (\boldsymbol\delta-\bar{\boldsymbol\delta}_w) \;\;\;   (\boldsymbol\delta-\bar{\boldsymbol\delta}_w)^\top \Bb (\boldsymbol\delta-\bar{\boldsymbol\delta}_w)}}.
\label{IB}
\end{eqnarray}

Defining $I_B$ on the basis of $\tilde{I}$ rather than upon the traditional formulation of the Moran's $I$ index, as given by Expression~(\ref{moran2}), allows $I_B$ to have the needed bounded fixed range and hence behaving properly as a spatial balance index.   

\section{Specification of the spatial weights matrix for $I_B$}

As it is commonly known, a natural specification of the $\Wb$ matrix does not exist and, therefore, the proper neighbourhood structure of units needs to be identified for the specific empirical circumstance under consideration. In the context of $I_B$, which refers to the spatial distribution of the sample inclusion indicator $\boldsymbol\delta$, it should be important to take both the distance among population units and their inclusion probabilities into account in order to identify the neighbourhood relationships. Let $0<\pi_i\leqslant1$ be the sample inclusion probability of population unit $i$. If $i$ were selected in the sample drawn from the population then $i$ would represent $1/\pi_i$ units in the population and, as a consequence, it would only be natural to consider that $i$ has $k_i=(1/\pi_i-1)$ neighbours in the population. The $k_i$ neighbours can be the $nearest$ neighbours of $i$ according to the distance. Let $\lfloor k_i \rfloor$ and $\lceil k_i \rceil$ be the inferior and superior integers of $k_i$, respectively. Let also $N_{\lfloor k_i \rfloor}$ be the set of the $\lfloor k_i \rfloor$ nearest neighbours of $i$, where $N_{\lfloor k_i \rfloor}=(N-1)$ if $1/\pi_i>N$. A proper criterion to specify $\Wb$ can then be

\begin{eqnarray}
w_{ij} = \begin{cases} 1 & \mbox{ if unit } j \in N_{\lfloor k_i \rfloor} \\ k_i-\lfloor k_i \rfloor & \mbox{ if unit } j \mbox{ is the } \lceil k_i \rceil \mbox{-th nearest neighbour of } i\\ 0 & \mbox{ otherwise.} \end{cases}
\label{wij}
\end{eqnarray}

\begin{figure}[htb!]
\begin{center}
\caption{Examples of stylized situations where $I_B$ reaches its extreme values. The locations of the sample units are represented with solid circles and the locations of the non-sampled units are represented with empty circles}
\includegraphics[scale=0.38]{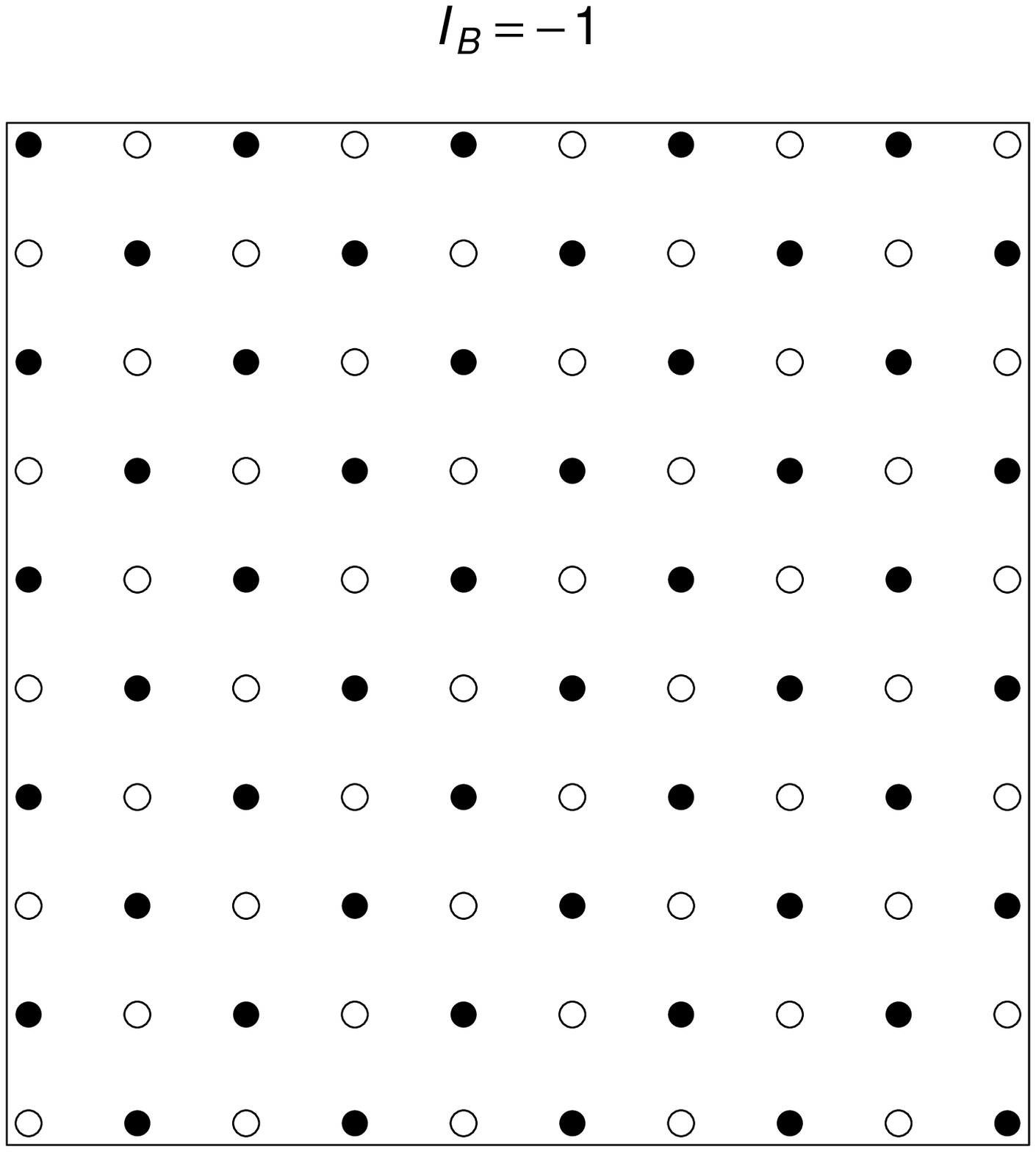}\includegraphics[scale=0.38]{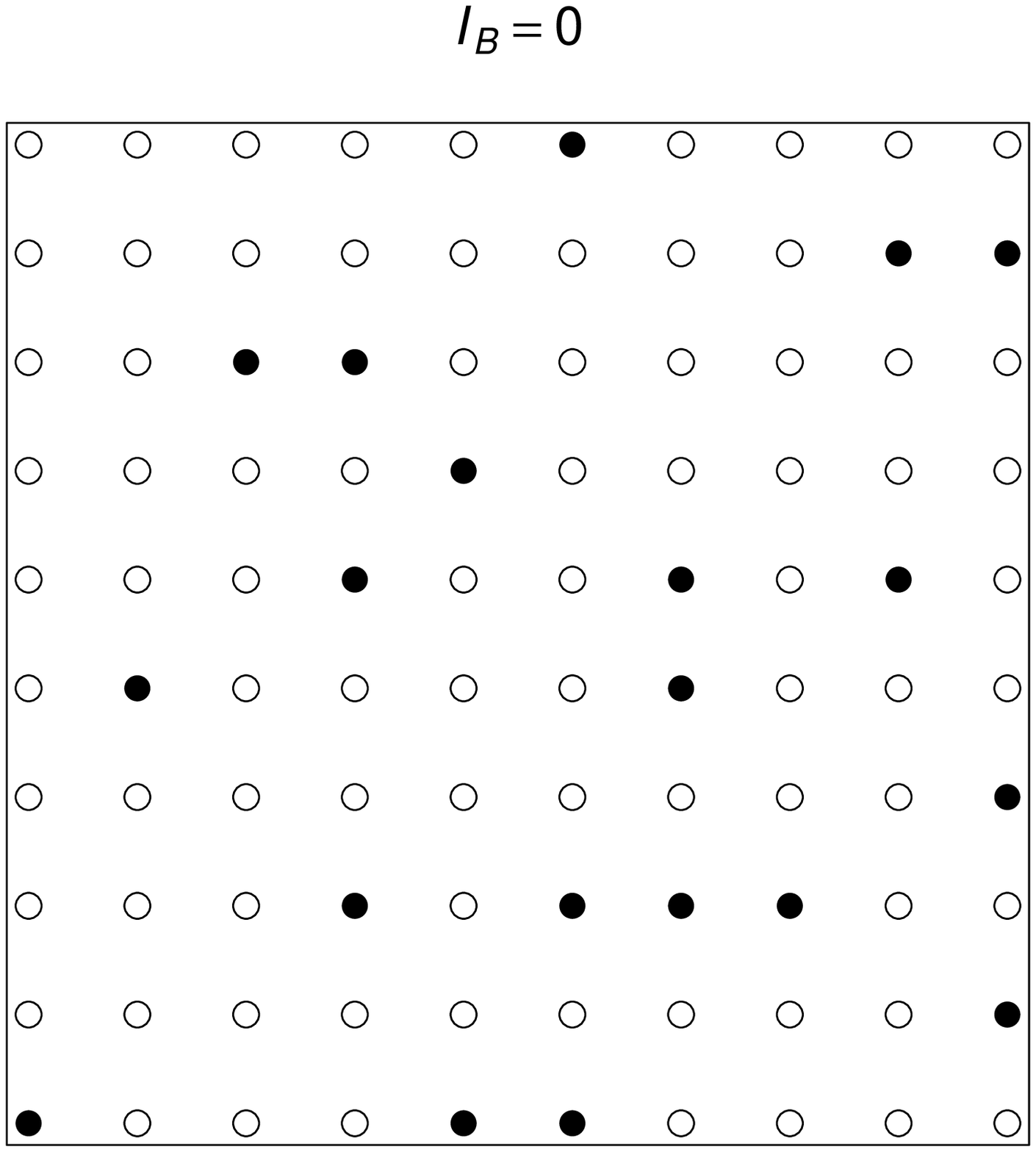}
\includegraphics[scale=0.38]{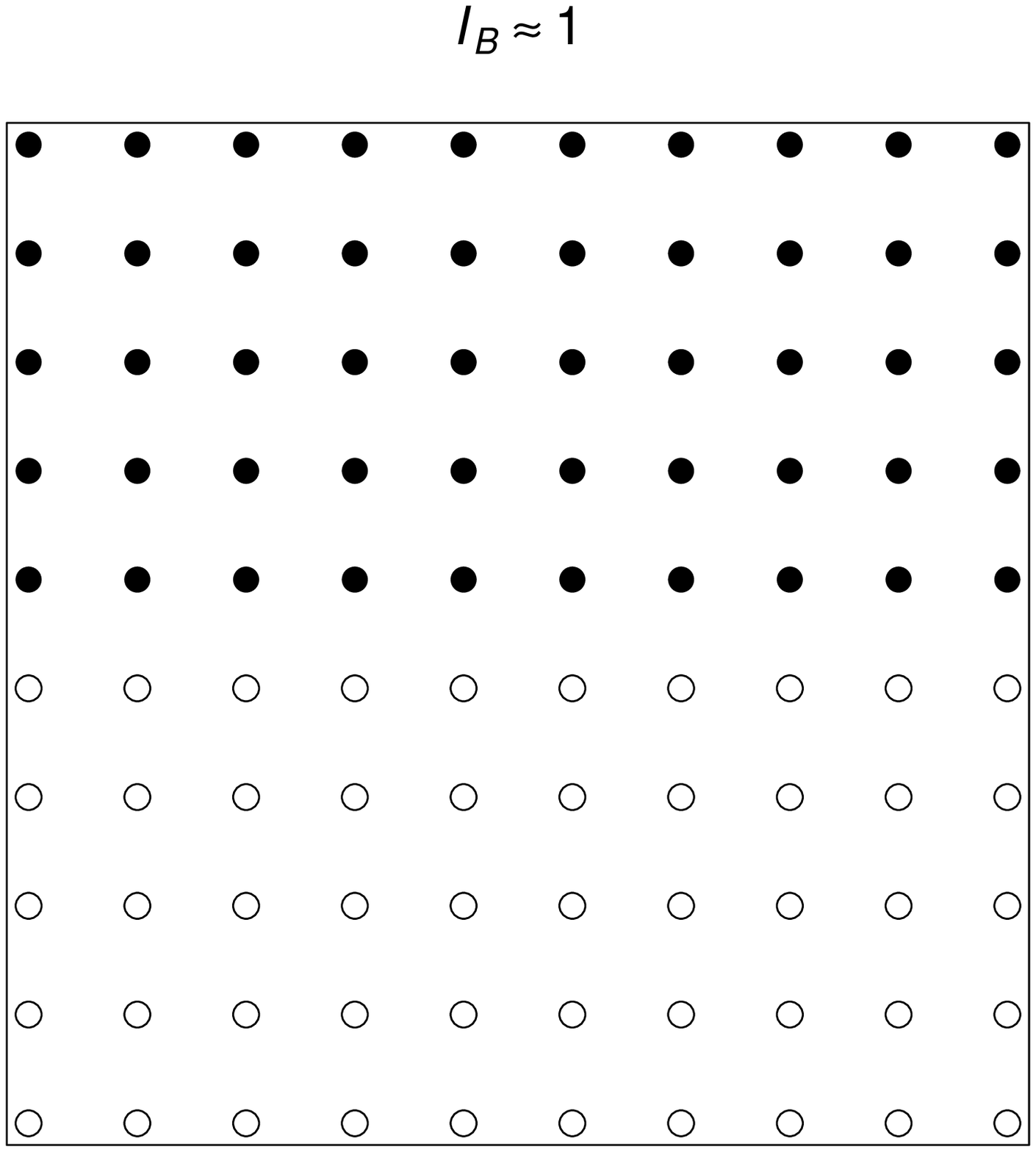}
\label{fig:graphIBbounds}
\end{center}
\end{figure}

This criterion implies, for example, that if $k_i=5.7$ then the 5 nearest neighbours of unit $i$ have a spatial weight of 1 while its 6-th nearest neighbour has a spatial weight of 0.7. In case there are two or more $\lceil k_i \rceil$-th units that have the same distance to $i$, we suggest to divide $w_{ij}$ equally among them.\\
As a way of illustration, Figure~(\ref{fig:graphIBbounds}) shows theoretical stylized situations where, with $w_{ij}$ specified according to Expression~(\ref{wij}), the use of $I_B$ achieves its upper bound of maximum spatial balance and its lower bound of maximum spatial clustering.   

\section{A simulation study based on artificial populations}

\begin{figure}[htb!]
\begin{center}
\caption{Maps of three target populations representing three illustrative situations that may occur in practical applications.}
\includegraphics[scale=0.35]{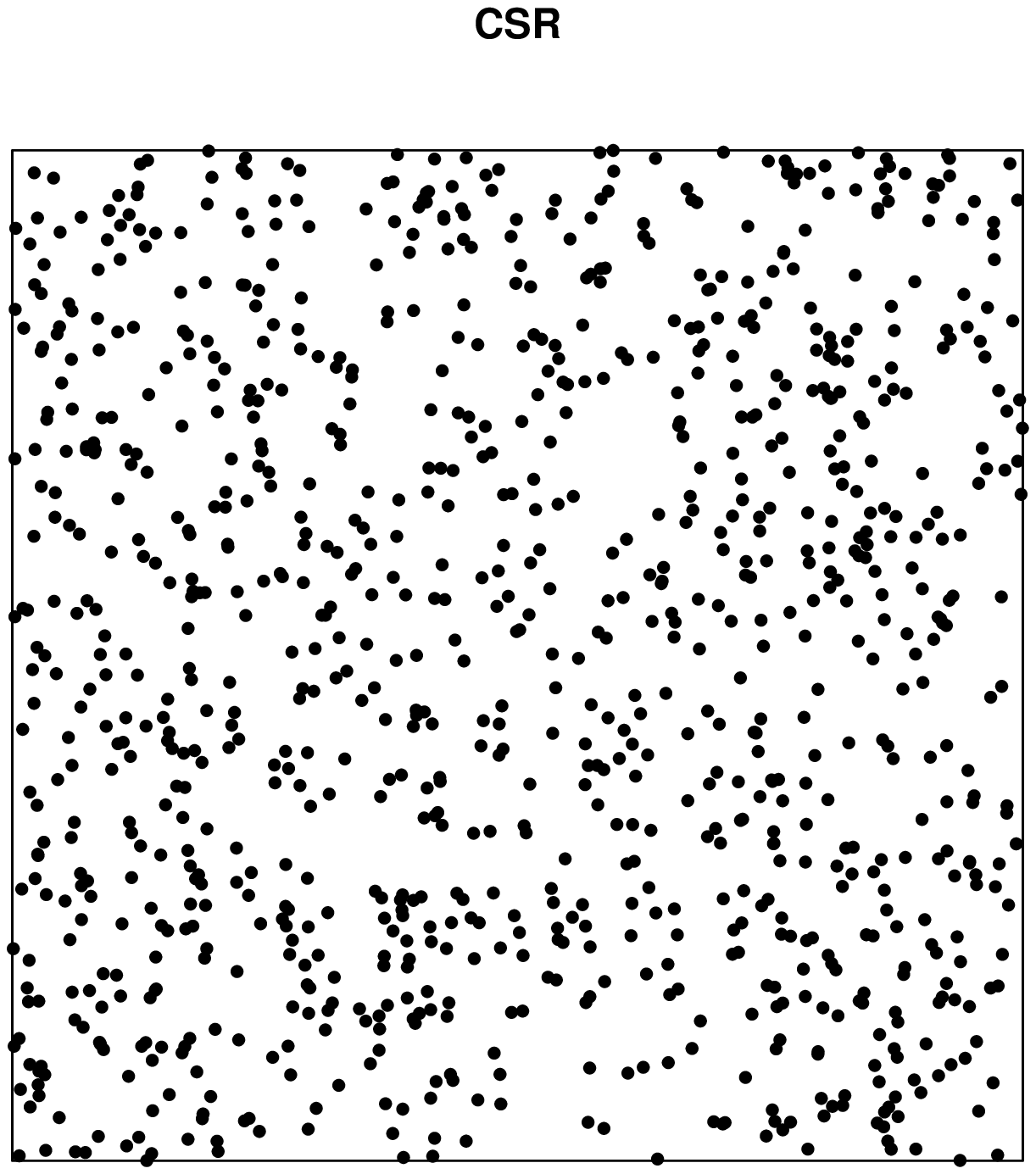}\includegraphics[scale=0.35]{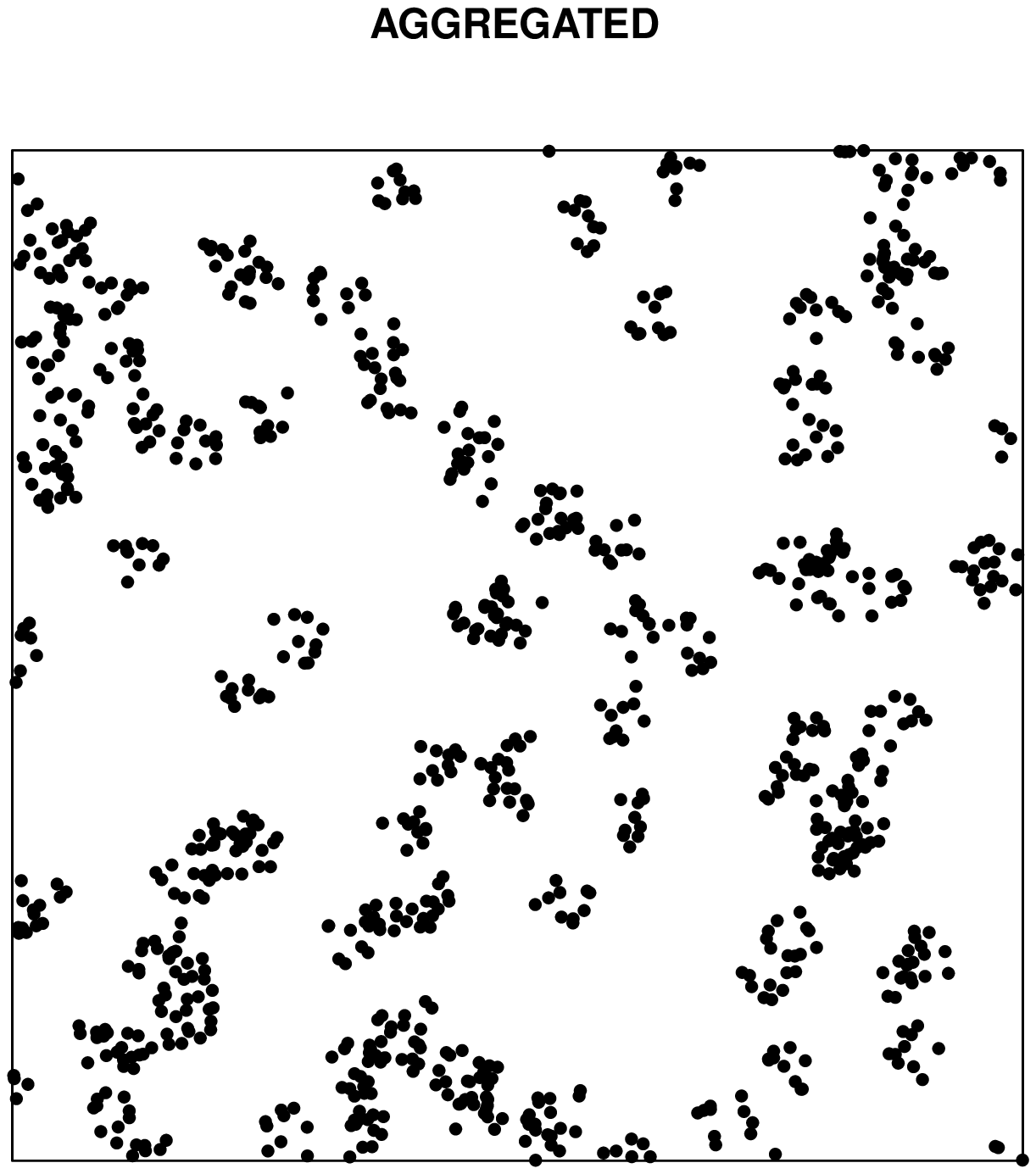}
\includegraphics[scale=0.35]{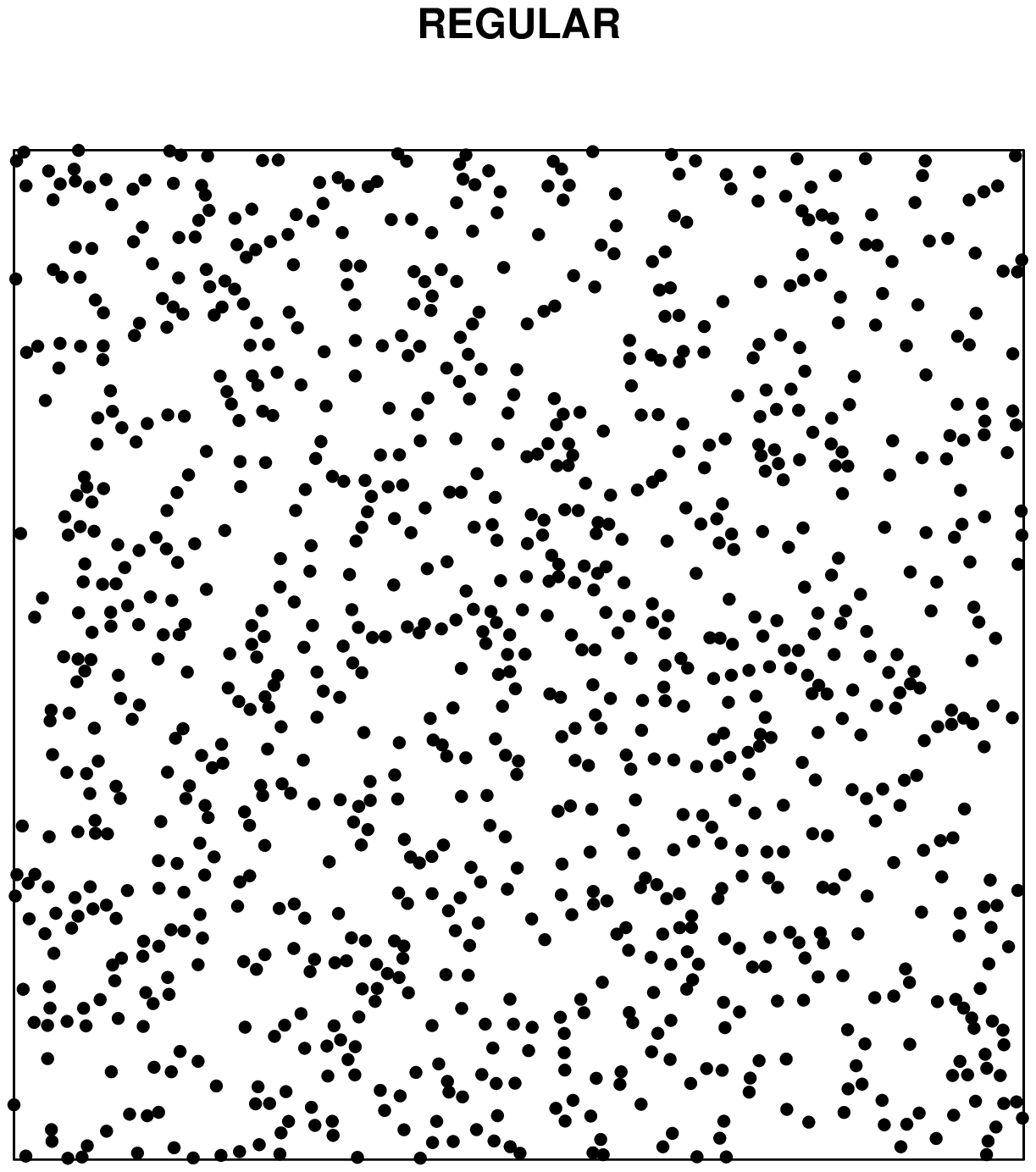}
\label{fig:MapsArtifPops}
\end{center}
\end{figure}

In order to assess the performance of $I_B$, a simple simulation experiment is carried out with the aim of studying how the index behaves according to different spatial characteristics of the surveyed population.
Three different target populations are generated representing the three main typical empirical circumstances. In particular, the following three artificial datasets of size $N = 1000$ are taken into consideration (see Figure~\ref{fig:MapsArtifPops}).

\begin{itemize}
\item Complete Spatial Randomness (CSR), which is a complete spatial random distribution of units on a planar unitary square. The units are generated according to a conditional homogeneous Poisson point process \cite[]{Diggle2013} with intensity parameter equal to 1000. As a consequence, they are homogeneously and independently located in space.
\item Aggregated sampling (AGGREGATED), which is a spatial clustered distribution of units on a planar unitary square. The units are generated according to a conditional Neyman-Scott point process \cite[]{NeymanScott1958} with 100 clusters and 10 points per cluster independently and uniformly distributed in a circular disc of radius 0.03 around their cluster centres. As a consequence, the units tend to be located close to each other leading to the occurrence of clusters of units.
\item Regular process (REGULAR), which is a spatial regular distribution of units on a planar square of size $1.5\times1.5$. The units are generated according to a Matern Model I process \cite[]{Matern1986} with an inhibition distance equal to 0.015. As a consequence, they tend to be located maintaining a certain distance to each other leading to the occurrence of a regular pattern.
\end{itemize}

Secondly, from each population (CSR, AGGREGATED or REGULAR), we select samples of size $n = 50, 100$ and $200$, respectively, with equal inclusion probabilities, $\pi_i = n/N$, and according to different sampling designs. In particular, the following sampling designs are compared.

\begin{itemize}
\item Local Pivotal Method Sampling (LPM), which produces spatially balanced samples. It selects units with the local pivotal method \cite[]{LPM2011}.
\item Simple Random Sampling (SRS), which produces samples without considering the spatial distribution of units. It selects units with simple random sampling without replacement.
\item Two-stage $k$-Clustered Sampling ($k$CLUST), which produces spatially clustered samples. It first divides the space into a $5\times5$ regular grid of cells. Secondly, it randomly selects $k$ cells. Finally, it randomly selects $n$ units among those located in the  $k$ cells.   
\end{itemize}

For all combinations of target population and sample size, a total of 10,000 samples are selected by each sampling design. For each selected sample, $B$, $I_M$ (i.e. the traditional non-standardized version of the Moran's $I$ index on the sample inclusion indicator) and $I_B$ are computed. To compare the performance of $I_B$ with that of the other indexes, we need to compare their average value computed over the 10,000 repeated samples. The results are presented in Tables \ref{tab:CSR}, \ref{tab:AGGREGATED} and \ref{tab:REG}.

\begin{table}[ht!]
\centering
\caption{Simulation results for population CSR. The values are means over all simulated samples.}
\label{tab:CSR}
\begin{tabular}{l r r r r r}
\hline
 & \multicolumn{3}{c}{Sampling design} \\
 Index & $k$CLUST & SRS & LPM \\
\hline
\multicolumn{4}{c}{$n=50$} \\
$B$ & 1.661 & 0.341 & 0.085 \\
$I_M$ & 0.093 & -0.001 & -0.04 \\
$I_B$ & 0.257 & -0.006 & -0.251 \\

\multicolumn{4}{c}{$n=100$} \\
$B$ & 2.588 & 0.351 & 0.104 \\
$I_M$ & 0.228 & -0.002 & -0.083 \\
$I_B$ & 0.417 & -0.006 & -0.339 \\

\multicolumn{4}{c}{$n=200$} \\
$B$ & 4.775 & 0.364 & 0.122 \\
$I_M$ & 0.559 & -0.001 & -0.184 \\
$I_B$ & 0.695 & -0.003 & -0.464 \\

\hline
\end{tabular}
\end{table}

\begin{table}[ht!]
\centering
\caption{Simulation results for population AGGREGATED. The values are means over all simulated samples.}
\label{tab:AGGREGATED}
\begin{tabular}{l r r r r r}
\hline
 & \multicolumn{3}{c}{Sampling design} \\
 Index & $k$CLUST & SRS & LPM \\
\hline
\multicolumn{4}{c}{$n=50$} \\
$B$ & 2.103 & 0.480 & 0.126 \\
$I_M$ & 0.103 & -0.001 & -0.041 \\
$I_B$ & 0.267 & -0.007 & -0.294 \\

\multicolumn{4}{c}{$n=100$} \\
$B$ & 3.645 & 0.503 & 0.119 \\
$I_M$ & 0.248 & -0.001 & -0.087 \\
$I_B$ & 0.432 & -0.005 & -0.405 \\

\multicolumn{4}{c}{$n=200$} \\
$B$ & 7.841 & 0.489 & 0.137 \\
$I_M$ & 0.896 & -0.001 & -0.180 \\
$I_B$ & 0.945 & -0.003 & -0.465 \\

\hline
\end{tabular}
\end{table}

\begin{table}[ht!]
\centering
\caption{Simulation results for population REGULAR. The values are means over all simulated samples.}
\label{tab:REG}
\begin{tabular}{l r r r r r}
\hline
 & \multicolumn{3}{c}{Sampling design} \\
 Index & $k$CLUST & SRS & LPM \\
\hline
\multicolumn{4}{c}{$n=50$} \\
$B$ & 1.540 & 0.325 & 0.077 \\
$I_M$ & 0.096 & -0.001 & -0.040 \\
$I_B$ & 0.263 & -0.006 & -0.245 \\

\multicolumn{4}{c}{$n=100$} \\
$B$ & 2.584 & 0.324 & 0.087 \\
$I_M$ & 0.229 & -0.001 & -0.083 \\
$I_B$ & 0.420 & -0.005 & -0.333 \\

\multicolumn{4}{c}{$n=200$} \\
$B$ & 4.406 & 0.330 & 0.111 \\
$I_M$ & 0.553 & -0.001 & -0.182 \\
$I_B$ & 0.691 & -0.003 & -0.442 \\

\hline
\end{tabular}
\end{table}

The index $I_B$ of spatial balance seems to behave properly regardless the characteristics of the surveyed population, the sample size and the sampling design. Indeed, in all simulations, $I_B$ follows the same trend of the $B$ index, that is it takes high values for the sampling design that produces spatially clustered samples and takes low values for the sampling design that produces spatially balanced samples. Moreover, it tends to +1 as the propensity to clustering of the sampling design increases, it tends to $-1$ as the tendency to spreading of the sampling design decreases and it equals approximately zero in the context of simple random sampling. When the degree of spatial balance is large, the non-standardized index $I_M$ is unable to measure the level of spreading properly because its range is not fixed.

\section{A simulation study based on a population of real business units}

\begin{figure}[htb!]
\begin{center}
\caption{Spatial distribution of the population of single-plant manufacturing firms in the province of Siena (Italy) in 2014. The size of circles is proportional to the number of employees.}
\includegraphics[scale=0.4]{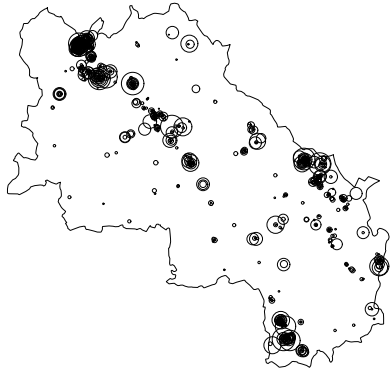}
\label{fig:mapSiena}
\end{center}
\end{figure}

To study also how $I_B$ performs with real data and with the use of unequal inclusion probabilities, we conduct a simulation experiment with a real population of 687 single-plant manufacturing firms, operating in the province of Siena (Italy) in 2014 (see Figure \ref{fig:mapSiena}). The dataset is a subset of the Statistical Register of Active Enterprises (ASIA) collected, managed and updated by the official Italian Statistical Institute (ISTAT). We refer to unequal inclusion probabilities proportional to an auxiliary size variable $v$ according to
$$
\pi_i = \min\left(1,\frac{C v_i}{\sum_{i=1}^{N} v_i}\right),
$$
where $C$ is chosen such that $\sum_{i\in U} \pi_i=n$.
The auxiliary variable $v$ is given by the number of employees.

This experiment considers three different sample sizes, say $n = 50$, $100$ and $150$, and two different unequal probability sampling designs, that is the maximum entropy sampling method (uMES) \cite[]{tille2006} and the local pivotal method (uLPM) both with unequal inclusion probabilities. Like SRS, uMES is a design that does not take into account the spatial distribution of units and hence selects samples that are not necessarily well spread in space. On the other hand, as we noted in Section~5, uLPM is specifically designed to select samples that are as spatially balanced as possible.
In all the six simulation scenarios 10,000 samples are selected and the means of $B$, $I_M$ and $I_B$ are computed over the repeated samples. The results for this experiment are found in Table \ref{tab:Siena} and $I_B$ is the best measure to assess the degree of spatial spreading of a sample because it reacts in the same way as $B$ to changes in the level of spatial balance but it has a more useful and interpretable measurement scale. Indeed, unlike $B$, $I_B$ can signals clearly how much a sample is spatially balanced or how much a sample is spatially clustered.

\begin{table}[ht!]
\centering
\caption{Simulation results for the population of single-plant manufacturing firms in the province of Siena (Italy) in 2014, with inclusion probabilities proportional to the number of employees. The values are means over all simulated samples.}
\label{tab:Siena}
\begin{tabular}{l r r}
\hline
 & \multicolumn{2}{c}{Sampling design} \\
 Index & uMES & uLPM \\
\hline
\multicolumn{3}{c}{$n=50$} \\
$B$ & 0.576 & 0.200 \\
$I_M$ & 0.002 & -0.014 \\
$I_B$ & 0.006 & -0.159 \\

\multicolumn{3}{c}{$n=100$} \\
$B$ & 0.533 & 0.184 \\
$I_M$ & 0.006 & -0.022 \\
$I_B$ & 0.018 & -0.150 \\

\multicolumn{3}{c}{$n=150$} \\
$B$ & 0.455 & 0.168 \\
$I_M$ & 0.007 & -0.030 \\
$I_B$ & 0.021 & -0.150 \\

\hline
\end{tabular}
\end{table}

Both simulation experiments have also provided indications about the practical achievable bounds of $I_B$. While the limits of the measure, namely $-1$ and $+1$, are achieved under rather extreme empirical circumstances, in real practical applications a value of $I_B$ around $-0.2$ already indicates a fairly high level of spatial balance.

\section{Conclusion}

In this paper, we introduced a new measure of the degree of spatial spreading of a sample using an \textit{ad hoc} normalized version of the Moran's $I$ index of spatial autocorrelation. The performance of the measure has been examined through both a simulation study on artificial spatial populations and an application to a population of real business units. The results have shown that the new measure has important advantages, when compared to the main existing measure based on the Voronoi polygons, regardless the spatial characteristics of the surveyed population, the sample size, the sampling design and the use of equal or unequal inclusion probabilities. In particular, it has a fixed and more interpretable scale of measurement with a specific benchmark value that allows to discriminates clearly, and in absolute terms, between absence and presence of spatial balance. 




\section*{Acknowledgements}

The authors would like to thank two anonymous reviewers for their valuable critical
remarks and suggestions that greatly contributed to improving the quality of the paper. They are also grateful to Danila Filipponi, Simonetta Cozzi and Patrizia Cella of the Italian National Institute of Statistics (Istat) who, according to a protocol between Istat and the University of Trento, have provided the data used in Section~6.

\section*{References}
\bibliographystyle{elsarticle-harv}
\bibliography{biblioMoran}




%%\bibitem[ ()]{}

%%\end{thebibliography}
\end{document}